\documentclass[aps,preprint,preprintnumbers,amsmath,amssymb]{revtex4}
\usepackage{amssymb}
\usepackage{graphicx}
\usepackage{dcolumn}
\usepackage{bm}
\usepackage{color}
\usepackage[breaklinks=true,colorlinks=true,linkcolor=blue,urlcolor=blue,citecolor=blue]{hyperref}

\makeatletter

\newcommand{\Rmnum}[1]{\expandafter\@slowromancap\romannumeral #1@}

\begin{document}

\title{Anomalous electronic structure and magnetoresistance in TaAs$_2$}

\author{Yongkang Luo\footnote[1]{Electronic address: ykluo@lanl.gov}, R. D. McDonald, P. F. S. Rosa, B. Scott, N. Wakeham, N. J. Ghimire\footnote[2]{Present address: Materials Science Division, Argonne National Laboratory, Argonne, Illinois 60439, USA}, E. D. Bauer, J. D. Thompson\footnote[3]{Electronic address: jdt@lanl.gov}, and F. Ronning\footnote[4]{Electronic address: fronning@lanl.gov}}

\address{$^1$Los Alamos National Laboratory, Los Alamos, New Mexico 87545, USA.}

\date{\today}


\maketitle

\textbf{The change in resistance of a material in a magnetic field reflects its electronic state.
In metals with weakly- or non-interacting electrons, the resistance typically increases upon the application of a magnetic field. In contrast, negative magnetoresistance may appear under some circumstances, \emph{e.g.}, in metals with anisotropic Fermi surfaces or with spin-disorder scattering and semimetals with Dirac or Weyl electronic structures. Here we show that the non-magnetic semimetal TaAs$_2$ possesses a very large negative magnetoresistance ($-$98\% in a field of 3 T at low temperatures), with an unknown scattering mechanism. Density functional calculations find that TaAs$_2$ is a new topological semimetal [$\mathbb{Z}_2$ invariant (0;111)] without Dirac dispersion, demonstrating that a negative magnetoresistance in non-magnetic semimetals cannot be attributed uniquely to the Adler-Bell-Jackiw chiral anomaly of bulk Dirac/Weyl fermions.
}\\

Magneto-transport continues to be an exciting topic in condensed matter physics. Some famous examples include discovering and understanding giant/collosal magnetoresistance\cite{Baibich-GMR,Binash-GMR,Tokura-CMR}, integer and fractional quantum Hall effects\cite{Klitzing-QHE,Tsui-FQHE}, Shubnikov-de Haas oscillations\cite{SdH}, and weak localization\cite{Lee-WL}. In metals with weakly- or non-interacting electrons, the resistance typically increases upon the application of a magnetic field due to the bending of electron trajectories\cite{Pippard-MR}. Negative magnetoresistance (MR) is observed in certain circumstances. To exploit these phenomena in applications it is essential to understand the scattering mechanisms involved. Low-carrier-density systems offer an interesting platform to explore the fundamental physics of scattering processes. A recent example is SrTiO$_{3-\delta}$ whose $T^2$-power law in resistivity, characteristic of a Landau Fermi liquid, cannot originate from simple electron-electron scattering, as often has been assumed\cite{Lin-T2Sci}. Semimetals can be considered as failed semiconductors with a negative indirect band gap. Consequently, these compensated systems, with approximately equal numbers of electrons and holes, have low effective masses due to the low band filling, which leads to rich magnetotransport phenomena including extremely large positive magnetoresistance (XMR) and ultrahigh mobilities exceeding those found in giant/collosal magnetoresistance systems\cite{Baibich-GMR,Binash-GMR,Tokura-CMR}. In addition, notions of topology have extended to semimetals as well. Accidental band crossings protected by symmetry allow electronic structures that are described by a massless Dirac equation. If either time reversal or inversion symmetry is broken, the four-fold (including spin) degenerate Dirac point splits into two Weyl points with opposite chirality. Typical examples are Cd$_3$As$_2$\cite{WangZ-Cd3As2DSM} and Na$_3$Bi\cite{WangZ-A3Bi} for Dirac semimetals, and $TmPn$ ($Tm$=Ta, Nb; $Pn$=As, P) for Weyl semimetals\cite{WengH-TmPn}. As a result of their exotic electronic structure, such semimetals host Fermi-arc surface states, XMR,  Shubnikov-de Haas (SdH) oscillations,  non-trivial Berry phases, and other related phenomena\cite{XuS-TaAsARPES,Lv-TaAsNP,Huang-TaAsLMR,Shekhar-NbP,Nirmal-NbAs,LuoY-NbAsSdH,HuJ-TaPLMR,XiongJ-Na3BiABJ}. Importantly, Dirac/Weyl semimetals are expected to have negative magnetoresistance when current is parallel to a magnetic field due to the Adler-Bell-Jackiw (ABJ) chiral anomaly mechanism\cite{Adler-ABJ, Bell-ABJ,Nielsen-ABJ,Son-ChirAnom}. The ABJ anomaly is a consequence of the chemical potential changing at each of the Weyl nodes, giving rise to an additional conduction channel, and has been taken as a smoking gun for the existence of a Dirac and/or Weyl semimetal.

If no accidental band crossings occur, can one still consider a semimetal as topologically non-trivial? The answer is yes. Similar to the classification for band insulators, $\mathbb{Z}_2$ topological indices ($\nu_0$;$\nu_1$$\nu_2$$\nu_3$) (strong and weak) are still appropriate for a regular semimetal due to the presence of a continuous energy gap between electron-like and hole-like bands. The surface states associated with weak (strong) topological indices are expected to be sensitive (immune) to disorder. Herein, we investigate a novel non-magnetic semimetal TaAs$_2$ that is homologous to the OsGe$_2$-type crystalline structure\cite{JeitschkoW-OsGe2} respecting inversion symmetry (Figure 1a). Magnetotransport measurements manifest a nearly compensated semimetal with low carrier density ($\sim$$10^{19}$ cm$^{-3}$), high mobility ($\sim$$10^{3}$ cm$^{2}$/Vs) and unsaturated XMR ($\sim$4,000,000\% at 65 T and 0.5 K). Further, angular dependent longitudinal magnetoresistance (LMR) measurements show pronounced negative MR ($\sim$$-$98\%), which suggests involvement of a ABJ chiral anomaly. Our first-principles calculations based on Density Functional Theory (DFT) confirm the semimetallicity of TaAs$_2$ but finds no evidence for a Dirac-like band-crossing. Instead, by computing the $\mathbb{Z}_2$ indices, (0;111), TaAs$_2$ is found to be a ``weak" topological material in all three reciprocal lattice directions but not a ``strong" topological material. Consequently, TaAs$_2$ should host surface states due to its electronic topology. We suggest that the very large negative magnetoresistance is a consequence of this novel topological state. Our observation of negative LMR in TaAs$_2$ also illustrates that the scattering mechanisms in (topological) semimetals are still not sufficiently understood.

\section*{Results}

Figure 1a shows the crystalline structure of TaAs$_2$. It crystallizes in a monoclinic structure with space group C12/m1 (No. 12, symmorphic). There are two chemical sites for As atoms in each unit cell, labeled As1 and As2, respectively. As1 and Ta form Ta-As planes. The interlayer coupling is bridged by As2 atoms, which reside near the central plane along the \textbf{c}-axis (see Figure 1b). Each Ta atom has eight nearest neighbors: five As1 and three As2. Figure 1c shows a TaAs$_2$ single crystal with a typical size on millimeter-grid paper. EDS analysis gives the mole ratio Ta:As=1:1.90(5), within experimental error consistent with the stoichiometric ratio. By XRD refinement, we deduce the crystalline lattice parameters listed in Table 1. Most importantly, inversion symmetry is respected in this compound.

In the absence of magnetic field, TaAs$_2$ shows a metallic Fermi-liquid-like $\rho_{xx}(T)$ profile, with a large residual resistivity ratio $RRR\equiv\rho_{xx}(300~\text{K})/\rho_{xx}(0.3~\text{K})$$\approx$100 (inset to Figure 2a), manifesting good sample quality. There is no signature of superconductivity above 0.3 K. When a magnetic field is applied, $\rho_{xx}(T)$ turns up and exhibits insulating-like behavior before it levels off at low temperature. Similar behavior is observed in other semimetallic materials \cite{Tafti-LaSb,WangK-NbSb2,Ali-WTe2XMR}. The insulating-like behavior becomes more and more pronounced as field increases, which leads to an XMR at low temperature. In Figure 2b, we show $MR(B)$[$\equiv$$(R(B)$$-$$R(0))/R(0)$$\times100\%$] measured at 0.5 K and in fields up to 65 T. The $MR$ reaches $\sim$4,000,000\% ($\sim$200,000\%) at 65 T (9 T), without any signature of saturation. Unlike the linear or sub-linear $MR(B)$ observed in the Dirac semimetal Cd$_3$As$_2$\cite{LiangT-Cd3As2} and the Weyl semimetals $TmPn$\cite{Huang-TaAsLMR,Shekhar-NbP,LuoY-NbAsSdH,HuJ-TaPLMR}, here $MR(B)$ generally obeys a parabolic field dependence (inset to Figure 2c), although the exponent decreases slightly at very high field (inset to Figure 2b). Such behavior is reminiscent of WTe$_2$\cite{Ali-WTe2XMR}, a candidate type-\Rmnum{2} Weyl semimetal\cite{Soluyanov-TypeIIWSM}.

In Figure 2e-f, we present Hall effect data. For all temperatures measured, the field-dependent Hall resistivity $\rho_{yx}$ is strongly non-linear and changes from positive at low field to negative at high field. The non-linearity of $\rho_{yx}(B)$ is reflected further by the divergence between the Hall coefficients $R_H(0)$ and $R_H(9~\text{T})$ as shown in the inset to Figure 2f. Here, $R_H(9~\text{T})$ is defined by $\rho_{yx}/B$ at $B$=9 T, and $R_H(0)$ is the initial slope of $\rho_{yx}(B)$ near $B$=0. All these features are characteristic signatures of multi-band effects. Indeed, $\rho_{yx}(B)$ can be well fit to a two-band model,
\begin{equation}
\rho_{yx}(B)=\frac{B}{e}\frac{(n_h\mu_h^2-n_e\mu_e^2)+(n_h-n_e)\mu_e^2\mu_h^2B^2}{(n_h\mu_h+n_e\mu_e)^2+[(n_h-n_e)\mu_e\mu_hB]^2},
\label{Eq.1}
\end{equation}
where $n$ and $\mu$ are respectively carrier density and mobility, and the subscript $e$ (or $h$) denotes electron (or hole). A representative fit to $\rho_{yx}(B)$ at $T$=0.3 K is shown in the inset to Figure 2e, and from this fit we obtain $n_e$=1.4(2)$\times10^{19}$ cm$^{-3}$, $n_h$=1.0(1)$\times10^{19}$ cm$^{-3}$, $\mu_e$=1.9(2)$\times10^{3}$ cm$^2$/Vs, and $\mu_h$=2.5(2)$\times10^{3}$ cm$^2$/Vs. The carrier densities are close to those estimated from the analysis of SdH oscillations [see \emph{\textbf{Supplementary Information}} (\emph{\textbf{SI}}) \emph{\textbf{\Rmnum{2}}}]. The low carrier density confirms TaAs$_2$ to be a semimetal. Furthermore, the imbalance between $n_e$ and $n_h$ implies that it is not a perfectly compensated semimetal\cite{LuoY-WTe2Hall}.

One important feature of topological Dirac/Weyl semimetals is the so-called ABJ chiral anomaly\cite{Nielsen-ABJ,Son-ChirAnom}. The ABJ anomaly is a result of chiral symmetry breaking when $\textbf{B}\cdot\textbf{E}$ is finite. This gives rise to a charge-pumping effect between opposite Weyl nodes. An additional contribution to the total conductivity is generated, \emph{i.e.}, $\sigma_{\chi}$$\propto$$B^2$, observable as a negative LMR\cite{Son-ChirAnom,XiongJ-Na3BiABJ}. In Figure 3a, we present the $MR(B)$ at 2 K and various $\phi$ ($\phi$ is the angle between \textbf{B} and electrical current \textbf{I}). Indeed, we observe a striking negative LMR when $\phi$=0. The $MR$ reaches $-$98\% before it starts to turn up weakly at high field (Figure 3f), which we ascribe to a small angular mismatch (see below). The negative LMR also persists to high temperatures $T$$>$150 K (cf Figure 3b). Compared with the chiral-anomaly-induced negative LMRs observed in Dirac/Weyl semimetals, such as Na$_3$Bi\cite{XiongJ-Na3BiABJ} and $TmPn$\cite{Huang-TaAsLMR,HuJ-TaPLMR}, the one seen in TaAs$_2$ is bigger in magnitude and survives at much larger $\phi$ and higher $T$. For example, Figure 3c plots $\rho_{xx}$ measured at 1 T and 2 K as a function of $\phi$, and the angular dependent $MR$ is sketched in a polar plot in Figure 3d. Clearly, the negative LMR survives for $\phi$ as large as 30 \textordmasculine. Note that the cusp near $B$=0 is not overcome until $\phi$$>$45 \textordmasculine (Figure 3a). In contrast to other systems\cite{XiongJ-Na3BiABJ,Huang-TaAsLMR,HuJ-TaPLMR}, because $\rho_{xx}(B)$ increases as $B^2$ when \textbf{B}$\perp$\textbf{I}, the slow rate of increase in MR in the vicinity of zero field makes it more robust against angular mismatch. This also allows the negative LMR in the limit of $B$$\rightarrow$0. Taking only 2\% residual resistivity at 3 T and the total carrier density $n_t$(=$n_e$+$n_h$)=2.4$\times 10^{19}$ cm$^{-3}$, we estimate the average transport mobility $\overline{\mu_{\parallel}}$=1.0$\times10^7$ cm$^2$/Vs. Using the Fermi-surface parameters of the electron-pocket as an example~(see \emph{\textbf{SI \Rmnum{2}}}), we further calculate the Fermi velocity $v_F$=7.9$\times10^5$ m/s, and transport relaxation time $\tau$=4.8$\times10^{-10}$ s. This means that the carriers can travel a distance (viz. mean free path) $l$=0.4 mm without backward scattering. Such an anisotropic MR and field induced low-scattering state would apparently find applications in electronic/spintronic devices, but the scattering mechanism is an open question.


Figure 4a shows the band structure and density of states (DOS) calculated with spin-orbit coupling (SOC). The semimetallic character can be seen by the low DOS at the Fermi level and the presence of small electron- and hole-bands. Figure 4b shows the Fermi surface (FS) topology calculated with SOC. The FS of TaAs$_2$ mainly consists of one hole- and two electron-pockets. The electron-pockets, located off the symmetry plane, are almost elliptical. The hole-pocket encompasses the \textbf{M} point at (1/2,1/2,1/2) but is more anisotropic with two extra ``legs". The abnormal FS structure of the hole pocket also is reflected in the complicated SdH frequencies discussed in the \emph{\textbf{SI \Rmnum{2}}}. Two additional electron-like pockets with vanishingly small size are observed intersecting the top of the Brillouin zone. Without SOC, accidental band crossings do occur as shown in the \emph{\textbf{SI \Rmnum{3}}}, and they can be classified as type-\Rmnum{2} Dirac points\cite{Soluyanov-TypeIIWSM}. Upon adding SOC, however, these Dirac points become gapped, and a careful survey over the entire Brillouin zone reveals no accidental band crossings in the vicinity of the Fermi level. The possibility of a Weyl semimetal is in any event excluded due to the preservation of both time reversal and inversion symmetries.

\section*{Discussion}

Due to the continuous gap in the band structure, the $\mathbb{Z}_2$ indices can be computed. The presence of inversion symmetry allows us to compute the topological indices ($\nu_0$;$\nu_1$$\nu_2$$\nu_3$) based only on the parities of the occupied wave functions at time-reversal-invariant-momenta (TRIM)\cite{Fu-PRB2007}. The results are shown in Figure 4c. (Refer to \emph{\textbf{SI \Rmnum{4}}} for more details.) The unoccupied states of the hole band at \textbf{M} do not influence the topological indices because these states have even parity. The product of parities over all the TRIM gives the value of the so-called ``strong" topological index $\nu_0$. As can be seen from Figure 4c, the electronic structure is trivial from this perspective. Nevertheless, all three ``weak" topological indices ($\nu_{1,2,3}$) are non-trivial. Hence, surface states are mandated by these weak topological indices, although they are believed to be sensitive to disorder.

We now return to the issue of the negative LMR. An electric current parallel to a magnetic field is not expected to experience a Lorentz force; however, in reality, negative LMR may exist stemming from a variety of mechanisms. First, because TaAs$_2$ is non-magnetic, a magnetic origin can be ruled out. Second, weak localization is also excluded, because $\rho_{xx}(T)$ conforms to Fermi-liquid behavior at low temperatures, and no $-\log T$ or any form of upturn signature can be identified. Third, negative LMR was also observed in materials such as PdCoO$_2$\cite{Kikugawa-PdCoO2LMR} with high FS anisotropy. To test the role of FS anisotropy, we measured the magnetoresistances of $R_{32,14}$ and $R_{14,32}$ with the schemes shown in the insets to Figure 3e. In the measurements of $R_{32,14}$, the current is parallel to \textbf{B}, and we derived a negative LMR, but the MR of $R_{14,32}$ is positive. Similar results were reproduced on several other samples with different shapes. Because the direction of current is arbitrary when referenced to the crystalline axes, these measurements imply that the observed negative LMR is locked to the relative angle between \textbf{E} and \textbf{B}, rather than pinned to particular FS axes. Fourth, an improperly made contact geometry may also cause negative LMR especially when the material shows a large transverse MR, known as the ``current-jetting" effect\cite{Pippard-MR,Yoshida-IJet,Ueda-IJet,Arnold-TaPLMR}. We have performed a series of careful LMR measurements with different contact geometries, and the results reveal that albeit a current-jetting effect can occur, the large negative LMR, however, is also intrinsic. \textbf{\emph{SI \Rmnum{5}}} provides more details.

To study further the features of this negative LMR, we plot $\Delta\sigma_{xx}(B)$=$\sigma_{xx}(B)$$-$$\sigma_{xx}(0)$ in the inset to Figure 3f. The low field part of $\Delta\sigma_{xx}$ can be well fitted to the form $C_3B^2$ (red line), which is consistent with an ABJ chiral conductivity $\sigma_{\chi}$. The absence of Dirac or Weyl points in TaAs$_2$, however, indicates that the negative LMR is not a consequence of the ABJ chiral anomaly as has been posited for other Dirac and Weyl semimetals. The fitting is converted back to $\rho_{xx}(B)$ as shown in the main frame of Figure 3f (red line). At high field, this fitting is gradually violated due to the emergence of a weak parabolic term in $\rho_{xx}(B)$, for which we successfully fitted $\Delta\sigma_{xx}(B)$ to the formula $C_1/(1+C_2B^2)$ (blue line). This weak positive MR is probably due to a small angular mismatch that causes a parabolic $MR(B)$ which becomes dominant as field strengthens (See also in \textbf{\emph{SI \Rmnum{5}}}).

Given the absence of alternative possibilities, an interesting question is whether the presence of topological surface states coexisting with a bulk semimetallic electronic structure could produce the large negative LMR as we observe. We note that conductivity corrections are found when surface states interact with bulk conduction states\cite{Garate}, although the observed effect here is an increase in the conductivity of a factor of $\sim$50. Having ruled out possible interpretations for the origin of a firmly established large, negative LMR in TaAs$_2$, this work calls for future theoretical and experimental work.

In summary, we find that single crystals of TaAs$_2$ grown by vapor transport are semimetals with extremely large, -unsaturating transverse magnetoresistance characteristic of high mobilities. Strikingly, TaAs$_2$ hosts a negative longitudinal magnetoresistance that reaches $-$98\%. TaAs$_2$ appears to be an example of a semimetal whose strong topological index is trivial, yet all three of its weak topological indicies are non-trivial. Similar properties also may exist in other OsGe$_2$-type $TmPn_2$ compounds where $Tm$=Ta and Nb, and $Pn$=P, As and Sb. As was the case for giant magnetoresistance, potential applications exist if the scattering mechanisms in these semimetals can be understood and manipulated. \\

\emph{Note added: When completing this manuscript, we became aware of several other related works\cite{WuD-TaAs2,LiY-TaSb2,WangY-TmAs2,YuanZ-TmAs2}.} \\

\section*{Methods}

\textbf{Sample synthesis and characterization}

Millimeter-sized single crystals of TaAs$_2$ were obtained as a by-product of growing TaAs by means of an Iodine-vapor transport technique with 0.05 g/cm$^3$ I$_2$. First, polycrystalline TaAs was prepared by heating stoichiometric amounts of Ta and As in an evacuated silica ampoule at 973 K for three days. Subsequently, the powder was loaded in a horizontal tube furnace in which the temperature of the hot zone was kept at 1123 K and that of the cold zone was $\sim$1023 K. Several TaAs$_2$ single crystals with apparent monoclinic shape were picked from the resultant and their monoclinic structure\cite{JeitschkoW-OsGe2} and stoichiometry were confirmed by x-ray diffraction (XRD) and energy dispersive x-ray spectroscopy (EDS). No I$_2$ doping was detected, and the stoichiometric ratio is fairly homogenous.

\textbf{Measurements}

Three TaAs$_2$ single crystals (labeled S1, S2 and S3) were polished into a plate with the normal perpendicular to the $\textbf{ab}$-plane. Ohmic contacts were prepared on the crystal in a Hall-bar geometry, and both in-plane electrical resistivity ($\rho_{xx}$) and Hall resistivity ($\rho_{yx}$, S1 only) were measured by slowly sweeping a DC magnetic field from $-$9 T to 9 T at a rate of 0.2 T/min. $\rho_{xx}$ ($\rho_{yx}$) was obtained as the symmetric (antisymmetric) component under magnetic field reversal. An AC-resistance bridge (LR-700) was used to perform these transport measurements in a 3-He refrigerator. Field-rotation measurements were carried out using a commercial rotator on a Physical Property Measurement System (PPMS-9, Quantum Design). Different contact geometries were made on S3 to show a possible current-jetting effect, and the measurements were performed in a 3-axis magnet. Magnetoresistance also was measured up to 65 T in a pulsed field magnet at the National High Magnetic Field Laboratory (NHMFL, Los Alamos). Several additional samples with different shapes were measured to confirm the reproducibility of negative LMR.

\textbf{DFT calculations}

Density functional theory calculations were performed using the generalized gradient approximation (GGA) as implemented in the WIEN2K code \cite{Wien2k} with the exchange correlation potential of Perdew-Burke-Ernzerhof (PBE)\cite{PBE}. Spin-orbit coupling on all atoms without relativistic local orbitals was included in a second variational scheme. The structure of TaAs$_2$ was obtained from Rietveld refinement (Table 1).

\emph{}\\

\textbf{Acknowledgments}\\
We thank Dmitrii Maslov and Hongchul Choi for insightful conversations. Sample preparation, and transport measurements were performed under the auspices of the Department of Energy, Office of Basic Energy Sciences, Division of Materials Science and Engineering. Electronic structure calculations were supported by the LANL LDRD program. Work at the NHMFL Pulsed Field Facility is supported through the National Science Foundation, the Department of Energy, and the State of Florida through NSF cooperative grant DMR-1157490. P. F. S. R acknowledges a Director's Postdoctoral Fellowship supported through the LANL LDRD program.

\textbf{Author contributions}\\
Y. L., J. D. T. and F. R. conceived and designed the experiments. N. J. G. and E. D. B. synthesized the samples. P. F. S. R. and B. S. characterized the crystals. Y. L. performed most of the measurements. R. D. M., N. W. and J. D. T. provided additional supporting measurements. F. R. carried out the first-principles calculations. Y. L., J. D. T. and F. R. discussed the data, interpreted the results, and wrote the paper with input from all the authors.

\textbf{Author information}\\
The authors declare no competing financial interests. Correspondence and requests for materials should be addressed to Y. Luo (ykluo@lanl.gov), J. D. Thompson (jdt@lanl.gov), or F. Ronning (fronning@lanl.gov).

\newpage
\textbf{Tables:}

\textbf{Table 1: Crystalline lattice parameters of TaAs$_2$.} Space group: C2/m1 (No. 12). $a$=9.370 \AA, $b$=3.394 \AA, $c$=7.771 \AA, $\alpha$=$\gamma$=90 \textordmasculine, and $\beta$=119.725 \textordmasculine. The atomic positions are:
\begin{table}[!h]
\begin{center}
\label{Tab.1}
\def\temptablewidth{1.0\columnwidth}
{\rule{\temptablewidth}{1pt}}
\begin{tabular*}{\temptablewidth}{@{\extracolsep{\fill}}ccccc}
Atoms                         & $x$         & $y$     & $z$       & Occ.    \\ \hline
Ta                            & 0.1574      & 0       & 0.1959    & 1.00    \\
As1                           & 0.4058      & 0       & 0.1072    & 1.00    \\
As2                           & 0.1394      & 0       & 0.5260    & 1.00    \\
\end{tabular*}
{\rule{\temptablewidth}{1pt}}
\end{center}
\end{table}\\

\newpage
\textbf{Figures:}

\begin{figure}[!h]
\includegraphics[width=10cm]{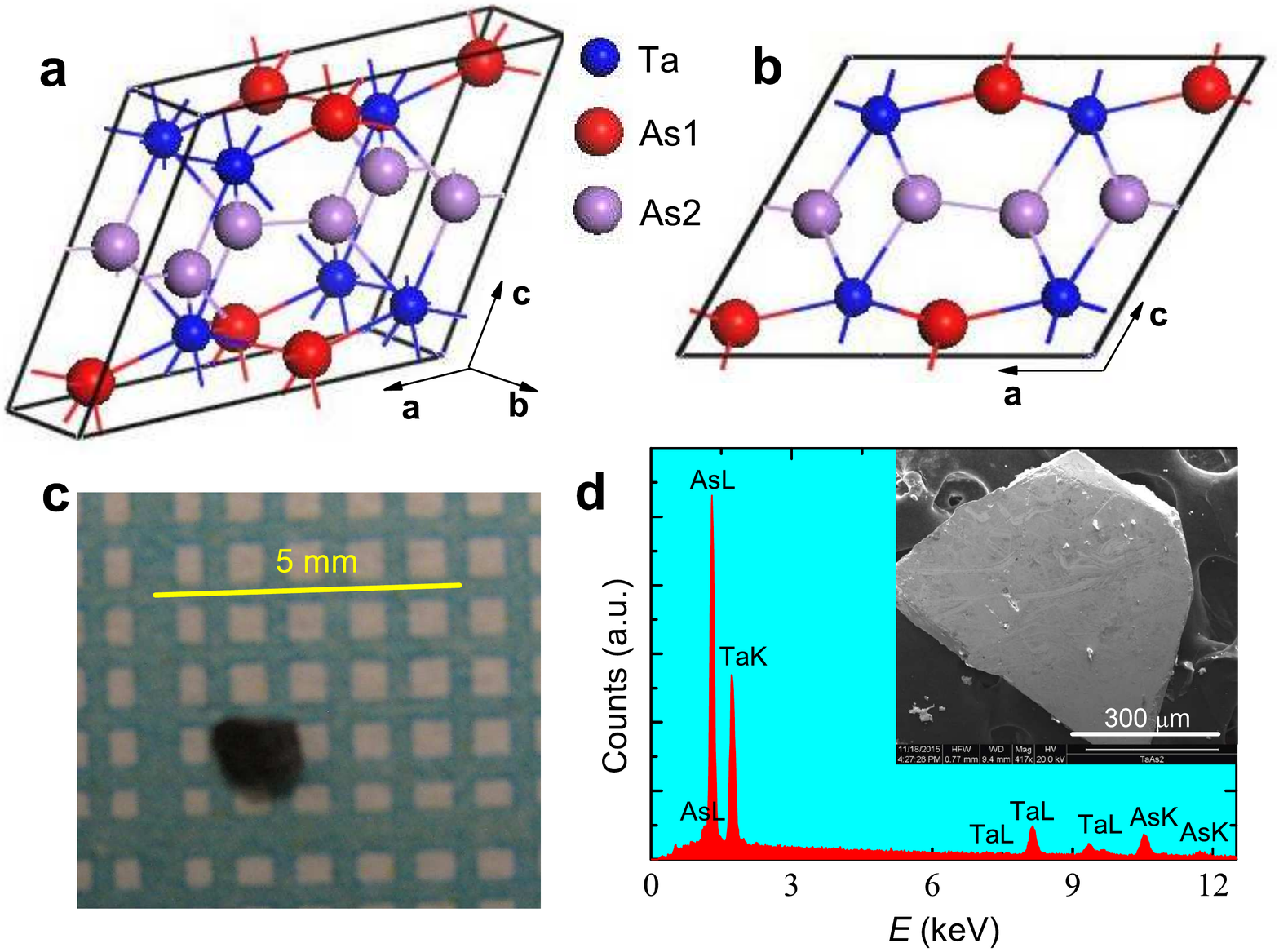}
\label{Fig1}
\end{figure}
\textbf{Figure 1 $|$ Crystalline structure of TaAs$_2$ and sample characterization.}
\textbf{a}, Crystalline structure of TaAs$_2$. \textbf{b}, A side view of TaAs$_{2}$ along (010)-axis. (c), A photograph of TaAs$_2$ single crystal on millimeter-grid paper. (d), A representative EDS spectrum of TaAs$_2$. The inset shows the SEM image of the same sample. \\
\newpage

\begin{figure}[!h]
\hspace*{-15pt}
\includegraphics[width=17cm]{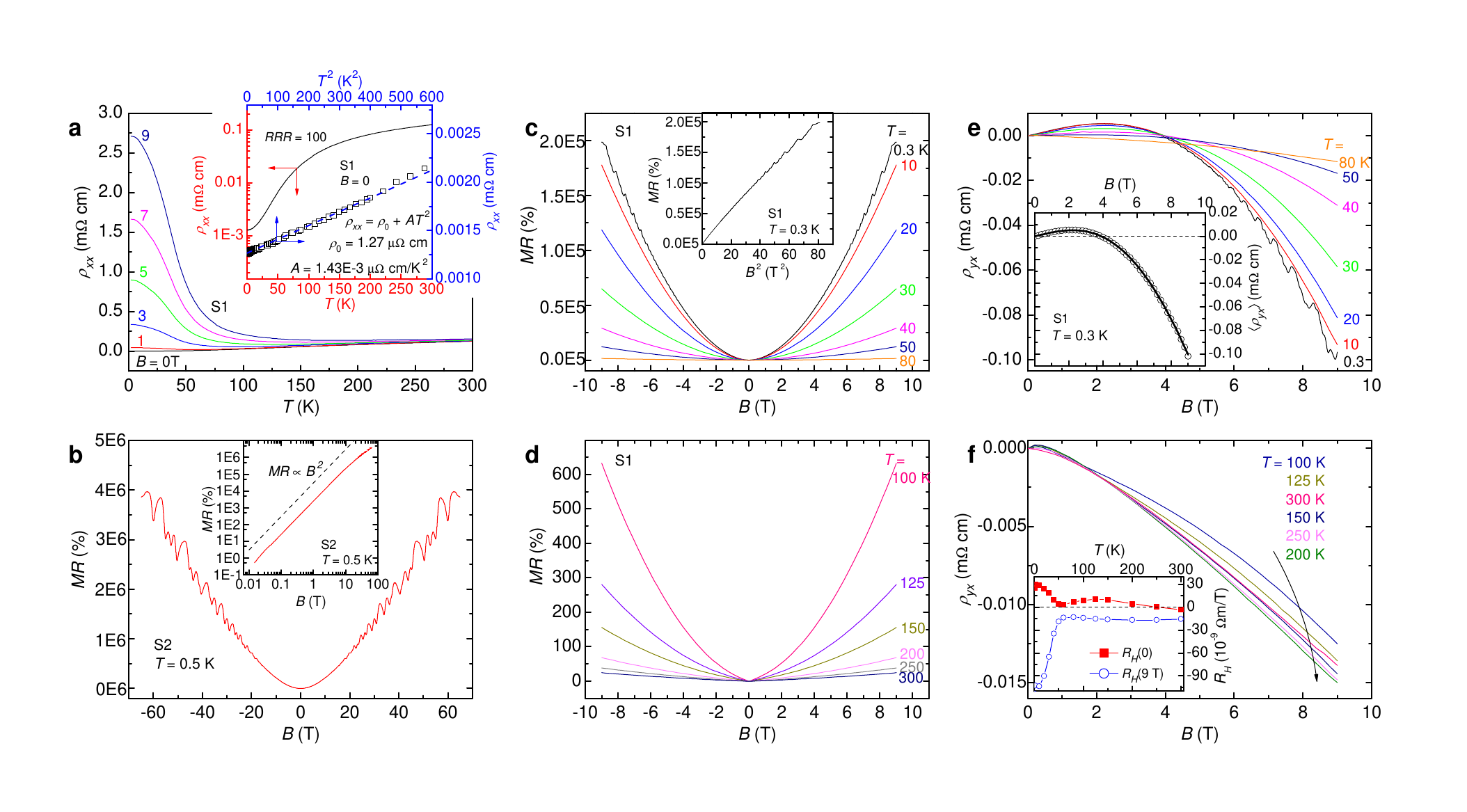}
\label{Fig2}
\end{figure}
\vspace{-25pt}
\textbf{Figure 2 $|$ Transport properties of TaAs$_2$.}
\textbf{a}, Temperature dependencies of $\rho_{xx}$ at selected magnetic fields. The inset shows $RRR$ and Fermi liquid behavior at $B$=0. \textbf{b}, Unsaturated $MR$ up to 65 T at 0.5 K. The inset demonstrates quadratic-like $MR(B)$. \textbf{c} and \textbf{d}, field dependent $MR$ at various temperatures. The inset to \textbf{c} shows $MR$ vs. $B^2$ at 0.3 K. \textbf{e} and \textbf{f}, Field dependent $\rho_{yx}$ at various temperatures. The inset to \textbf{e} displays a two-band fit of $\langle\rho_{yx}(B)\rangle$ at 0.3 K. The inset to \textbf{f} displays the Hall coefficient $R_H$ as a function of $T$.\\

\newpage

\begin{figure}[!h]
\vspace{-20pt}
\includegraphics[width=16cm]{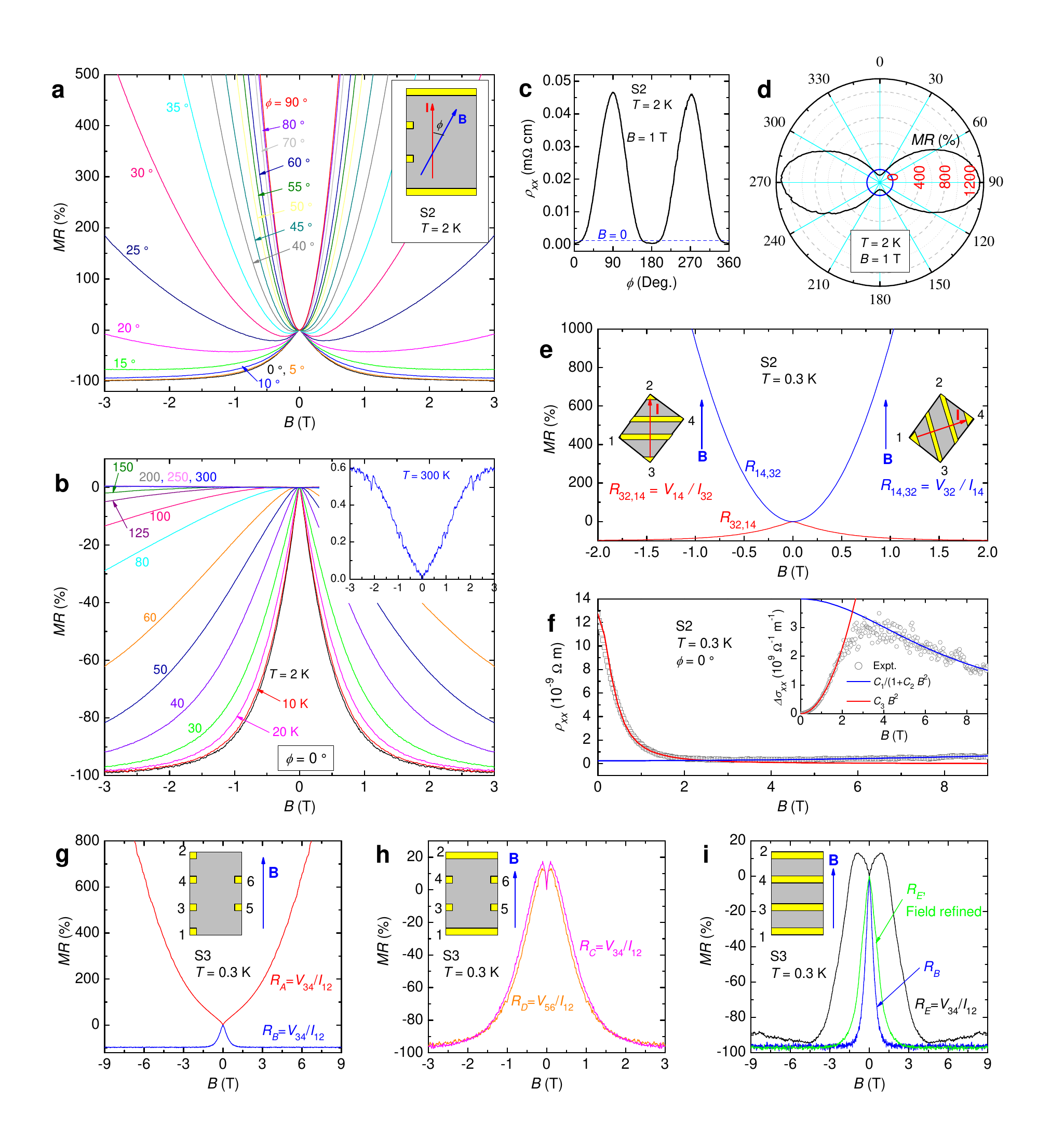}
\label{Fig3}
\end{figure}
\vspace{-20pt}
\textbf{Figure 3 $|$ Longitudinal magnetoresistance (LMR) of TaAs$_2$.}
\textbf{a}, Field-dependent $MR$ of TaAs$_2$ with various angles $\phi$ at 2 K. The inset shows the configuration of the measurements. \textbf{b}, $MR$ at different temperatures, measured at $\phi$=0. The inset displays the data at 300 K. \textbf{c}, The angular dependence of $\rho_{xx}$ at 2 K and 1 T. \textbf{d}, A polar plot of $MR$ at 2 K. \textbf{e}, MR with two different measurement geometries, $R_{32,14}$=$V_{14}/I_{32}$ (red) and $R_{14,32}$=$V_{32}/I_{14}$ (blue). Schematic sketches of the geometry are shown in the insets. \textbf{f}, Theoretical fits of $\rho_{xx}(B)$ and $\Delta\sigma_{xx}(B)$. The high-field part of $\Delta\sigma_{xx}(B)$ is fit to $C_1/(1+C_2B^2)$ (blue), and the low-field part is fit to $C_3B^2$ (red). The measurements were done in the contact geometry as shown in the inset to \textbf{a}. 

\newpage

\begin{figure}[!h]
\includegraphics[width=16cm]{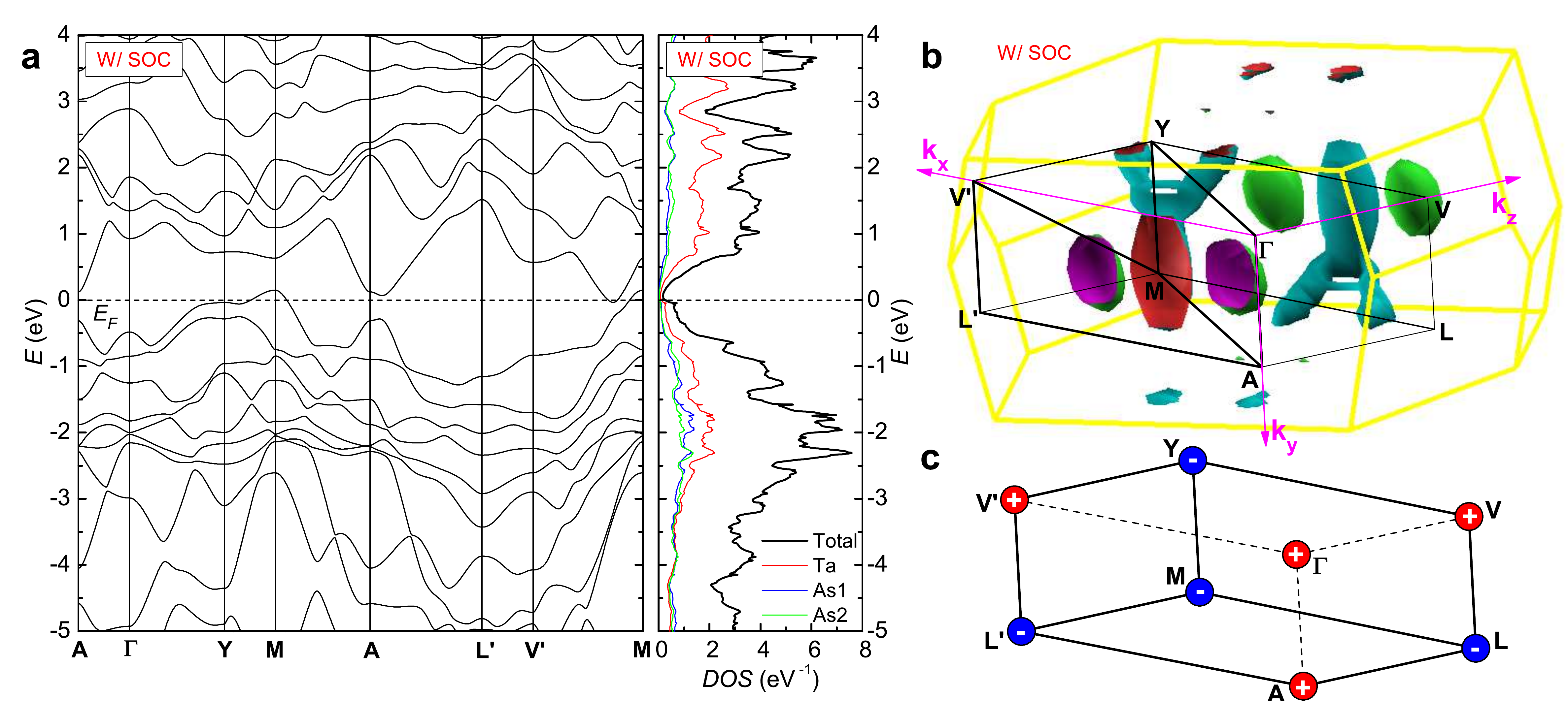}
\label{Fig4}
\end{figure}
\textbf{Figure 4 $|$ DFT calculations of TaAs$_2$ with SOC.}
\textbf{a}, Band structure and DOS of TaAs$_2$. \textbf{b}, FS topology and TRIM points. \textbf{c}, Parity of the TRIM at the monoclinic Brillouin zone. \\

\newpage

\vspace{-15pt}

\begin{center}
\emph{\textbf{Supplementary Information: } }\\
\textbf{Anomalous magnetoresistance in TaAs$_2$}\\
\end{center}

\vspace{10pt}

In this \emph{\textbf{Supplementray Information}} (\emph{\textbf{SI}}), we provide magnetization, Shubnikov-de Hass (SdH) quantum oscillations, additional Density-functional-theory (DFT) calculations, topological indices, and longitudinal magnetoresistance (LMR) that further support the discussion and conclusions of the main text.\\



\emph{\textbf{SI \Rmnum{1}: Magnetization of TaAs$_2$}}\\

\begin{figure}[!h]
\vspace*{-15pt}
\includegraphics[width=8.5cm]{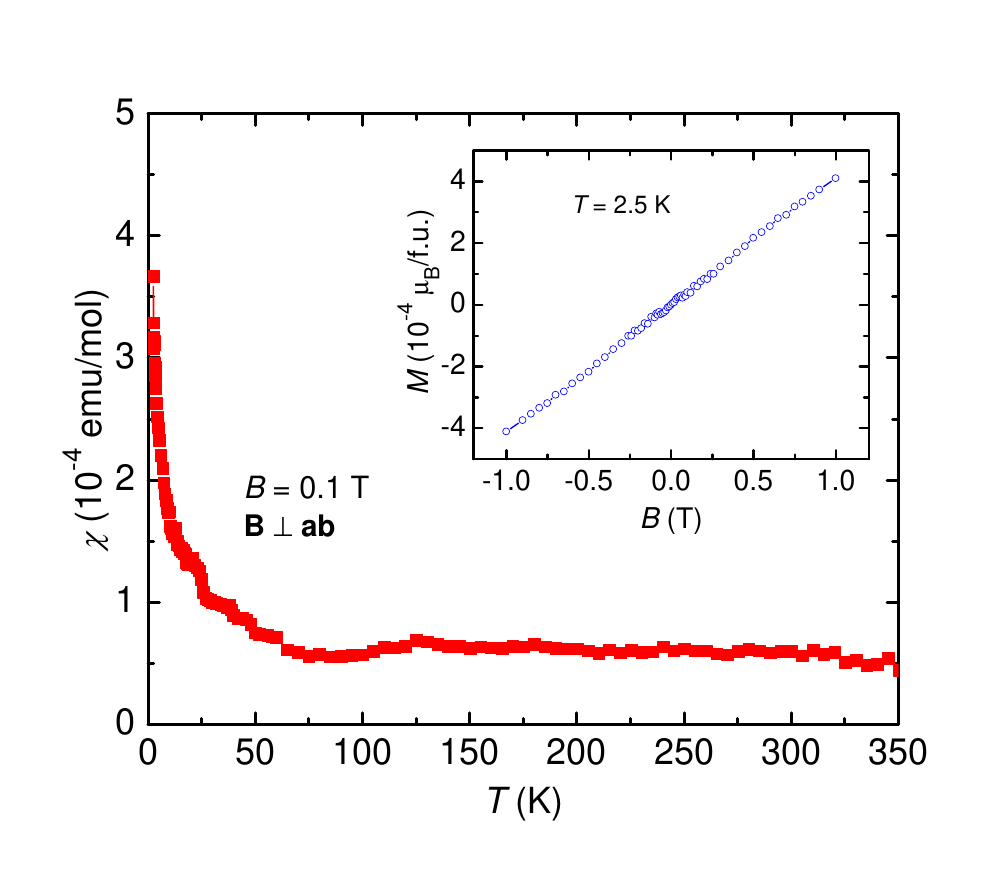}
\label{Fig.S1}
\end{figure}
\vspace*{-10pt}
\emph{\textbf{Figure S1 $|$ Magnetization of TaAs$_2$.} Main frame, temperature dependence of magnetic susceptibility $\chi$. The inset shows isothermal field dependent magnetization at 2.5 K. }\\

The magnetic properties of TaAs$_2$ are displayed in Figure S1. These measurements were taken on sample S1 with an external magnetic field $\textbf{B}$$\perp$$\textbf{ab}$, using a Quantum Design Magnetic Property Measurement System (MPMS-5). The main frame of Figure S1 shows the temperature dependence of magnetic susceptibility $\chi(T)$. The value of $\chi$ is 5.94$\times10^{-5}$ emu/mol at room temperature, and remains essentially unchanged down to 50 K. The weak upturn at low temperature is likely an impurity contribution. Such a Pauli-paramagnetic $\chi(T)$ curve can not be described by the well-known Curie-Weiss law, manifesting the absence of intrinsic local moments. This is further confirmed by the isothermal field dependent magnetization at 2.5 K as shown in the inset to Figure S1. The magnetization reaches only 4$\times10^{-4}$ $\mu_B$/f.u. at 1 T. All these results demonstrate that TaAs$_2$ is a non-magnetic compound, and the large transverse magnetoresistance and negative longitudinal magnetoresistance discussed in the main text do not have a magnetic origin. In addition, we may also conclude that time reversal symmetry is respected in TaAs$_2$.\\

\emph{\textbf{SI \Rmnum{2}: SdH oscillations}}\\

Another important feature of the magnetotransport property of TaAs$_2$ is the SdH quantum oscillation atop of the large magnetoresistance signal. The analysis of SdH data is somewhat complicated. Further systematic measurements under higher magnetic field are needed to better clarify the details of the Fermi surface (FS) topology. We leave this task for future work. To estimate the carrier density we analyse the SdH oscillations based on our angular measurements up to 9 T.

The SdH effect can be observed in both $\rho_{xx}(B)$ (Figure S2a) and $\rho_{yx}(B)$ (Figure S2b). We derive the oscillatory part from $\Delta\rho_{ij}$=$\rho_{ij}$$-$$\langle\rho_{ij}\rangle$ ($i$,$j$=$x$,$y$), where the non-oscillatory part $\langle\rho_{ij}\rangle$ is obtained by a fourth-order polynomial fit to $\rho_{ij}(B)$. The obtained $\Delta\rho_{xx}(B)$ and $\Delta\rho_{yx}(B)$ are displayed with the right axes of Figure S2a and S2b, respectively.

By taking the Fast Fourier Transformation (FFT) of $\Delta\rho_{xx}$ as a function of $1/B$, we obtain multiple SdH oscillation frequencies as shown in Figure S2c. The two fundamental frequencies are $F_{\alpha}$=104(2) T and $F_{\beta}$=130(2) T. The decaying amplitude of SdH oscillations with temperature is described by the Lifshitz-Kosevich (LK) formula\cite{SLuoYK-NbAsSdH}:
\begin{equation}
\frac{\Delta\rho_{xx}}{\langle\rho_{xx}\rangle}\propto\frac{2\pi^2k_BT/\hbar\omega_c}{\sinh(2\pi^2k_BT/\hbar\omega_c)},\tag{S1}
\label{Eq.S1}
\end{equation}
in which $\omega_c$=$\frac{eB}{m^*}$ is the cyclotron frequency with $m^*$ being the effective mass. We tracked the FFT amplitudes of $\alpha$- and $\beta$-orbits as a function of $T$ in Figure S2(d). Fitting these data points to the LK formula, we derived the effective masses, $m^*_{\alpha}$=0.083(1) $m_0$ and $m^*_{\beta}$=0.078(1) $m_0$, where $m_0$ is the mass of a free electron. These small effective masses are similar to those in other topological materials, {\it e.g.}, 0.089 $m_0$ for the 3D topological insulator Bi$_2$Te$_2$Se\cite{SRen-Bi2Te2Se,SOng-Bi2Te2Se}, 0.043 $m_0$ for the Dirac semimetal Cd$_3$As$_2$\cite{SWangJ-Cd3As2SdH} and 0.033-0.066 $m_0$ for the Weyl semimetal NbAs\cite{SLuoYK-NbAsSdH}, but are much smaller than those of NbSb$_2$ ($\sim$1 $m_0$), an iso-structural analog\cite{SWangK-NbSb2}, and for WTe$_2$ ($\sim$0.4 $m_0$)\cite{SCaiP-WTe2PreSdH}, a candidate type-II Weyl semimetal\cite{SSoluyanov-TypeIIWSM}.

\begin{figure}[!h]
\includegraphics[width=16cm]{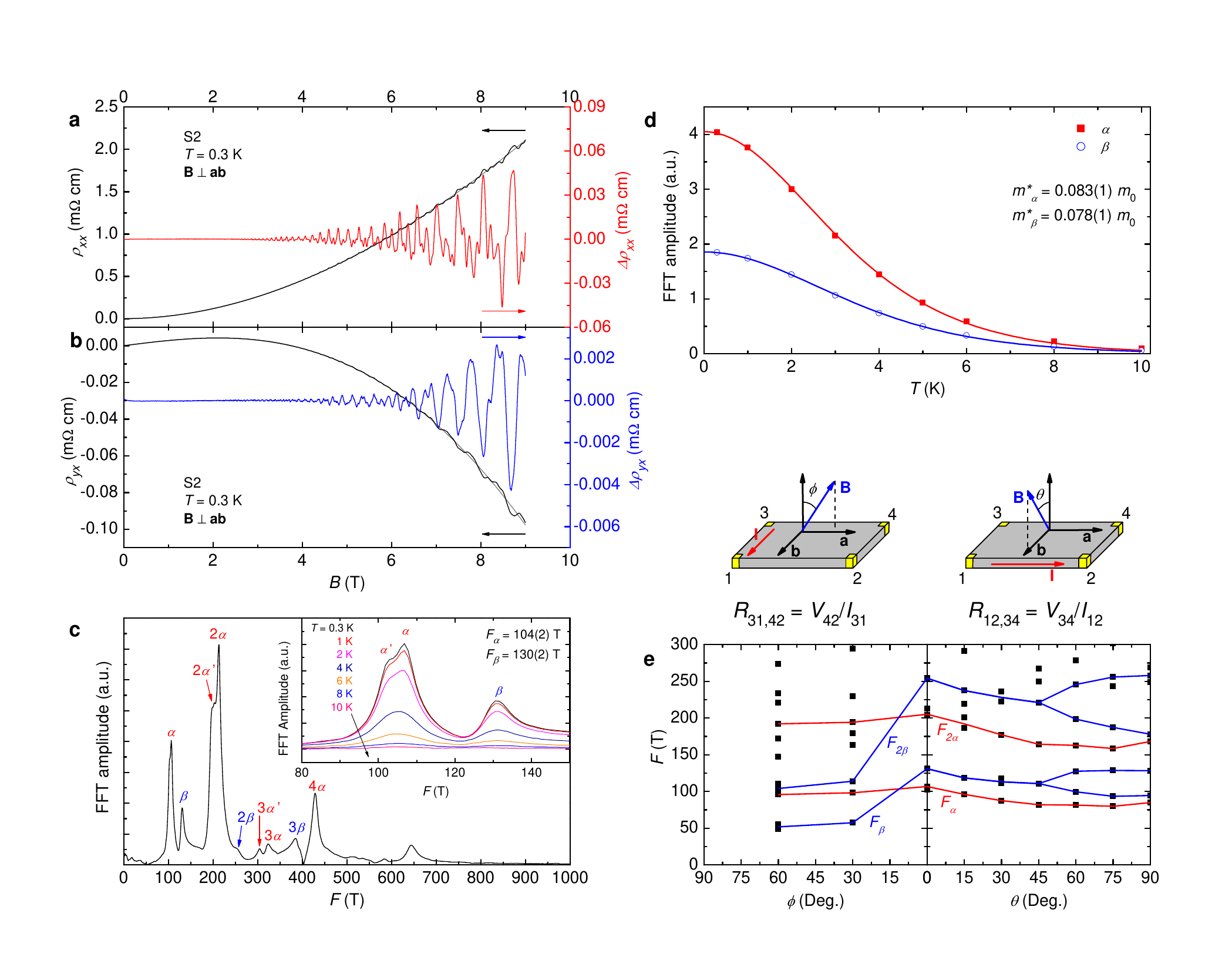}
\label{Fig.S2}
\end{figure}
\vspace*{-15pt}
\emph{\textbf{Figure S2 $|$ SdH quantum oscillations of TaAs$_2$.} \textbf{a} and \textbf{b}, Field dependence of $\rho_{xx}$ and $\rho_{yx}$ at 0.3 K for $\textbf{B}$$\perp$$\textbf{ab}$. The grey lines are the fourth-order polynomial fitting. The right axes show oscillatory contributions, $\Delta\rho_{xx}$ and $\Delta\rho_{yx}$, respectively. \textbf{c}, FFT spectrum of $\Delta\rho_{xx}(1/B)$. The inset is a zoom-in view at selected temperatures. \textbf{d}, Temperature dependent FFT amplitudes of $\alpha$- and $\beta$-pockets. Fitting to the LK formula results in effective masses $m_{\alpha}^*$=0.083(1) $m_0$, $m_{\beta}^*$=0.078(1) $m_0$. The solid lines are theoretical fittings to LK formula. \textbf{e}, Angular ($\theta$ and $\phi$) dependence of FS cross-sectional extrema. The solid lines are guidelines. The scheme of the measurements is also depicted.}\\

We also performed angular dependent SdH oscillation measurements. $\theta$ and $\phi$ respectively depict the angles between electrical current $\textbf{I}$ and magnetic field $\textbf{B}$ when the field is rotated in two different ways, see Figure S2e. The oscillatory frequencies are shown in Figure S2e, and the possible guidelines for their angular dependencies are also presented. Combining this with the DFT calculations addressed in the main text, we may assume the $\alpha$-orbit is due to an electron-pocket, and the $\beta$-orbit is due to a hole-pocket.

For a three dimensional system, the carrier density is correlated with the size of Fermi surface via,
\begin{equation}
n=\frac{gk_F^xk_F^yk_F^z}{3\pi^2}=\frac{g}{3\pi^2}\sqrt{\frac{8e^3F_xF_yF_z}{\hbar^3}},\tag{S2}
\label{Eq.S2}
\end{equation}
where $g$ is the multiplicity of the Fermi surface in the first Brillouin zone, $k_F^i$($i$=$x$,$y$,$z$) is the magnitude of Fermi momentum along $\textbf{i}$-axis, and $F_i$ is the oscillatory frequency with magnetic field $\textbf{B}$ parallel to $\textbf{i}$-axis. Here, we have included the spin degeneracy and adopted the Lifshitz-Onsager correlation $F$=$\frac{\hbar}{2\pi e}S_F$, in which $S_F$ is the extremal cross-sectional area of the Fermi surface. As a rough estimate, we treat the electron FS as a spherical pocket, with $F_{\alpha x}$=$F_{\alpha y}$=$F_{\alpha z}$$\approx$100 T and $g_e$=2. The carrier density of electrons is thus calculated $n_e$=1.1$\times10^{19}$ cm$^{-3}$. Due to the complex Fermi surface topology, we are not able to calculate the carrier density of holes directly, but considering the magnitude of oscillation frequency and the multiplicity $g_h$=1, it is reasonable to set $n_e$ as the upper limit of $n_h$. These estimates are quantitatively in agreement with the Hall effect measurement discussed in the main text.

Finally, it should be pointed out that the observed $F_{\alpha}$ peak weakly splits into two at low temperature, labeled as $\alpha$ and $\alpha'$ (cf the inset to Figure S2c). This is better seen in their higher-order harmonics. This splitting seems to disappear as field rotates. It is likely that this splitting is caused by a certain sub-structure on the FS which results in additional cross-sectional extrema at a particular angle. $F_{\beta}$ shows a much stronger angular dependence (Figure S2e); however, this might be not surprising considering its complicated topology. More systematic angular SdH measurements under higher magnetic field are required to further resolve this. \\

\emph{\textbf{SI \Rmnum{3}: Additional DFT calculations}}\\

In Figure S3, we present additional DFT calculation results. Figure S3a shows the band structure and density of states (DOS) calculated without spin-orbit coupling (SOC) along the same path as in Figure 4 of the main text where SOC is included. Two extra cuts, \textbf{K}-\textbf{M} and \textbf{K'}-\textbf{K"}, as indicated in Figure S3b, are presented in detail to illustrate the effects of SOC. A small gap of $\sim$60 meV between conduction- and valence-bands is observed along \textbf{K'}-\textbf{K"}. More interestingly, a band-crossing occurs along \textbf{K}-\textbf{M}. However, in contrast to regular Dirac/Weyl semimetals, the band-crossing appears at the contact of electron- and hole-pockets, rather than a point-like FS (viz. type-\Rmnum{1} Dirac/Weyl point). This reminds us of the type-\Rmnum{2} Dirac/Weyl semimetal proposed by Soluyanov {\it et al} recently\cite{SSoluyanov-TypeIIWSM}. However, this crossing is not protected by the symmetry of the crystal lattice. Hence, when SOC is included the Dirac-like point becomes gapped. The comparisons between band structures without and with SOC along the two cuts are shown in Figure S3c and S3d, respectively. When SOC is turned on, the band crossing along \textbf{K}-\textbf{M} opens a small gap with magnitude $\sim$117 meV. Intuitively, the band gap in \textbf{K'}-\textbf{K"} is further enlarged by SOC.

\begin{figure}[!h]
\includegraphics[width=16cm]{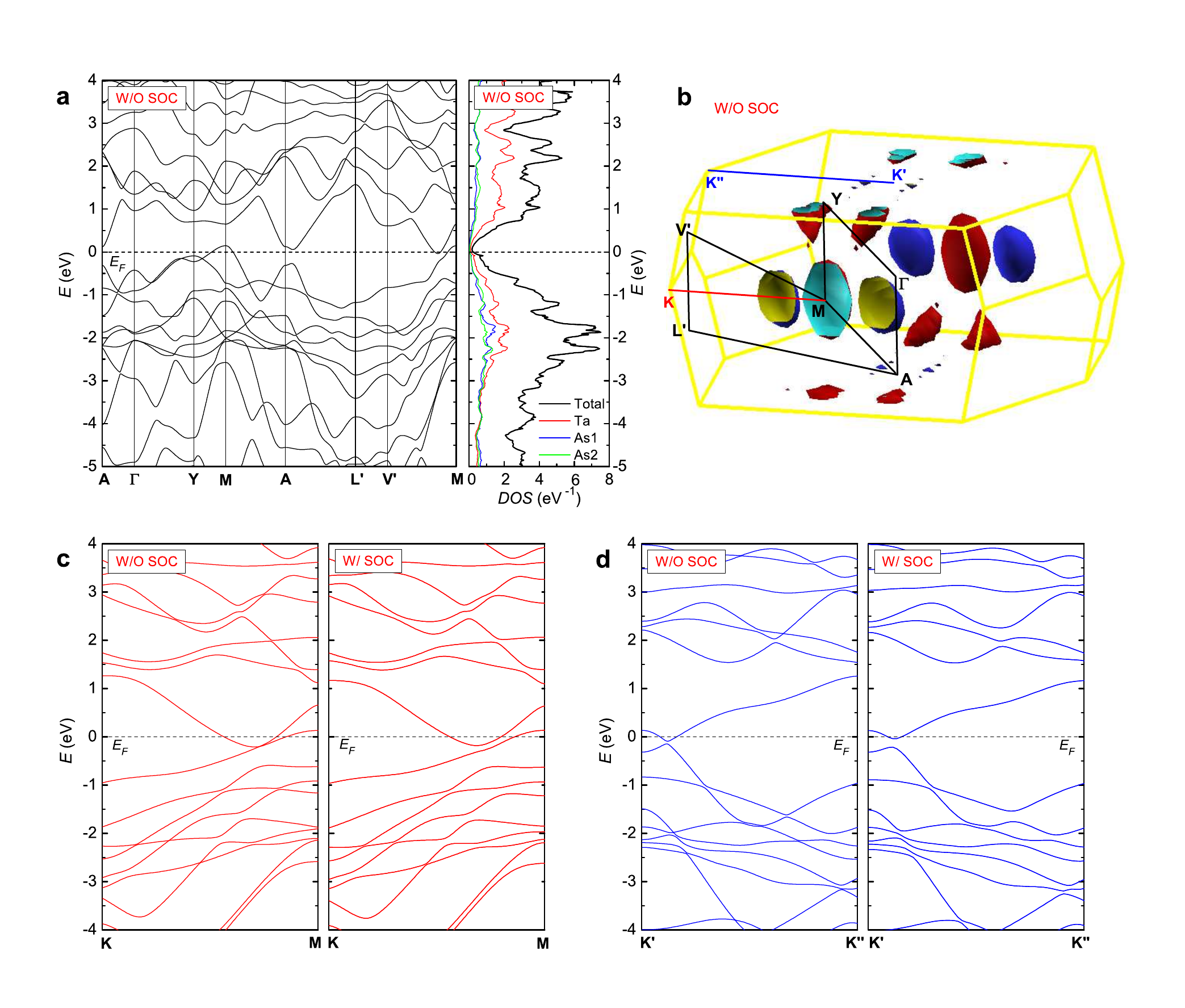}
\label{Fig.S3}
\end{figure}
\vspace*{-10pt}
\emph{\textbf{Figure S3 $|$ DFT calculations of TaAs$_2$ without SOC.} \textbf{a}, Band structure and DOS of TaAs$_2$ calculated without SOC. \textbf{b}, FS topology with the \textbf{k}-path cuts used in panels \textbf{c} and \textbf{d}. \textbf{c} and \textbf{d} respectively display the comparisons of band structure along \textbf{K}-\textbf{M} and \textbf{K'}-\textbf{K"} with and without SOC.}\\

\emph{\textbf{SI \Rmnum{4}: $\mathbb{Z}_2$ topological invariants}}\\

The DFT calculations also allow us to analyse the $\mathbb{Z}_2$ topological invariants ($\nu_0$;$\nu_1$$\nu_2$$\nu_3$) for a centrosymmetric crystal\cite{SFu-PRB2007}. Given the parity eigenvalue of the pair of occupied degenerate bands, we determine the parity ($\delta$) of the each time-reversal-invariant-momentum (TRIM) points in the BZ as summarized in Table S1 (see also in Figure 4c).\\

\emph{\textbf{Table S1: Parity of the TRIM of the monoclinic Brillouin zone.} Calculated based on DFT with SOC turned on.}
\begin{table}[!h]
\begin{center}
\label{Tab.S1}
\def\temptablewidth{1.0\columnwidth}
{\rule{\temptablewidth}{1pt}}
\begin{tabular*}{\temptablewidth}{@{\extracolsep{\fill}}ccc||ccc}
TRIM                      & ($k_x, k_y, k_z$)      & $\delta$   ~~~~~~~~~~~    & TRIM                    & ($k_x, k_y, k_z$)    & $\delta$       \\ \hline
$\mathbf{\Gamma}$         & (0, 0, 0)              & $+$1       ~~~~~~~~~~~    & $\mathbf{A}$            & (0, 1/2, 0)          & $+$1           \\
$\mathbf{Y}$              & (1/2, 0, 1/2)          & $-$1       ~~~~~~~~~~~    & $\mathbf{M}$            & (1/2, 1/2, 1/2)      & $-$1            \\
$\mathbf{V}$              & (0, 0, 1/2)            & $+$1       ~~~~~~~~~~~    & $\mathbf{L}$            & (0, 1/2, 1/2)        & $-$1           \\
$\mathbf{V'}$             & (1/2, 0, 0)            & $+$1       ~~~~~~~~~~~    & $\mathbf{L'}$           & (1/2, 1/2, 0)        & $-$1            \\
\end{tabular*}
{\rule{\temptablewidth}{1pt}}
\end{center}
\end{table}

A sign change of $\delta$ between two TRIM points manifests a band inversion. The strong topological index $\nu_0$ is defined by $(-1)^{\nu_0}$=$\prod\delta_i$, where $\prod$ goes through all the eight TRIM points. This leads to $\nu_0$=0. The weak topological indices are calculated via the product of $\delta_i$ at four coplanar TRIM points in the BZ\cite{SAutes-Bi4I4}, \emph{i.e.}, $(-1)^{\nu_1}$=$\delta_{\mathbf{M}}\delta_{\mathbf{Y}}\delta_{\mathbf{V'}}\delta_{\mathbf{L'}}$, $(-1)^{\nu_2}$=$\delta_{\mathbf{M}}\delta_{\mathbf{L'}}\delta_{\mathbf{A}}\delta_{\mathbf{L}}$, $(-1)^{\nu_3}$=$\delta_{\mathbf{M}}\delta_{\mathbf{L}}\delta_{\mathbf{V}}\delta_{\mathbf{Y}}$, and we obtain $\nu_1$=$\nu_2$=$\nu_3$=1. This analysis suggests that TaAs$_2$ is a ``weak" topological material in all three reciprocal lattice directions, but not a ``strong" topological material. \\

\emph{\textbf{SI \Rmnum{5}: Additional LMR}}\\

Improperly made contact geometry may also cause negative LMR, especially when the material shows a large transverse MR, a so-called ``current-jetting" effect\cite{SYoshida-IJet,SUeda-IJet,SArnold-TaPLMR}. To test if this will affect the LMR we have observed in TaAs$_2$, we performed a series of measurements with various contact geometries, and results are summarized in Figure S4. These measurements were done in a 3-axis magnet ($B_{x,y,z}$) to precisely tune the field orientation. We start with the case of inhomogeneous current as shown in Figure S4a, from which we indeed obtain significantly different MR behaviors in $R_A$=$V_{34}/I_{12}$ and $R_B$=$V_{56}/I_{12}$. Although $R_{B}$ decreases rapidly with increasing $B_z$, the MR of $R_A$ is dramatically positive. It might not be too surprising for such a divergence in the context of a current-jetting effect: due to the large transverse MR, the path 1-5-6-2 becomes more and more resistive under magnetic field, and therefore more and more current is accumulated along the path 1-3-4-2, which inevitably increases (decreases) the voltage drop between the electrodes 3 (5) and 4 (6).

\begin{figure}[!h]
\hspace*{-28pt}
\includegraphics[width=18cm]{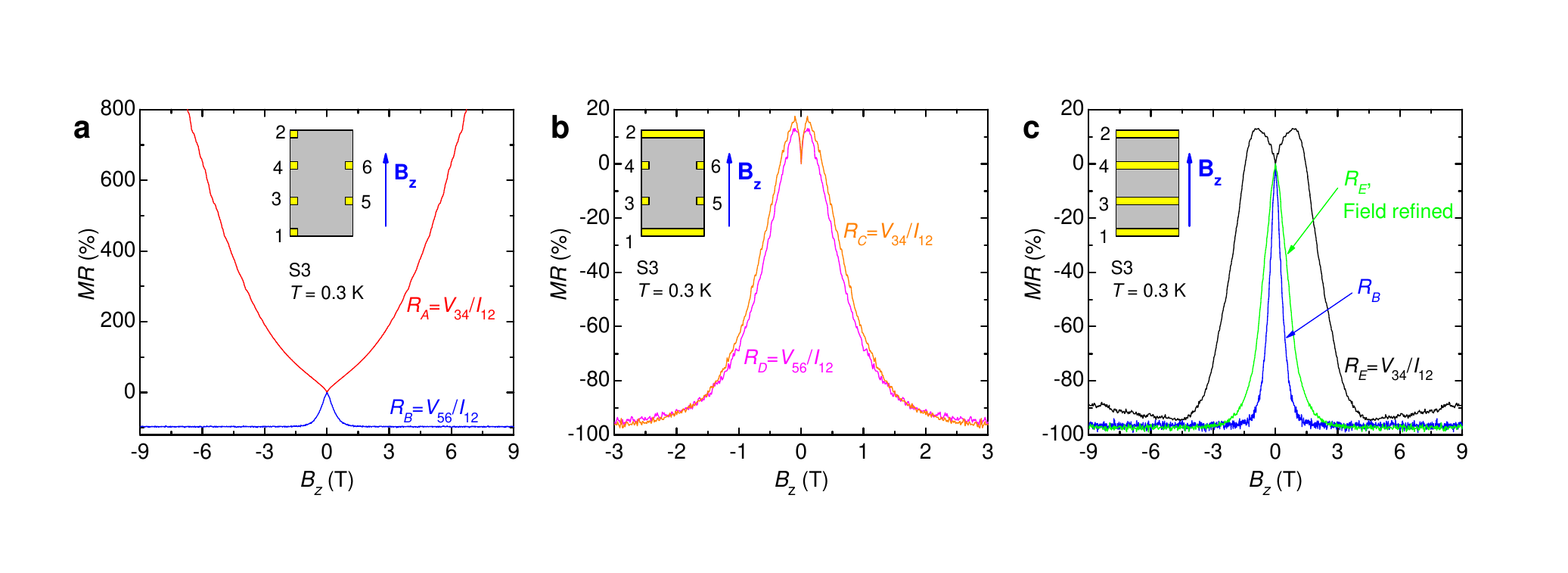}
\label{Fig.S4}
\end{figure}
\vspace*{-20pt}
\emph{\textbf{Figure S4 $|$ LMR measurements with various contact geometries.} \textbf{a}, An inhomogeneous current indeed causes different magnetoresistance behaviors in $R_A$=$V_{34}/I_{12}$ and $R_B$=$V_{56}/I_{12}$. \textbf{b}, Measurements in a Hall-bar geometry. Negative LMR is seen in both $R_C$ and $R_D$. \textbf{c}, A four-probe measurements ($R_E$) with all the contacts fully across the width. A small positive MR is seen on top of the large negative LMR. Intrinsic negative LMR can be seen after the field direction is carefully tuned.}\\

To avoid this current-jetting effect, we improved the current homogeneity by fully painting the current leads across the end faces of the sample (Figure S4b). The derived resistances are now labeled $R_C$ and $R_D$, respectively. Both of $R_C$ and $R_D$ exhibit negative LMR although in a small region near $B_z$=0 the MR initially turns up. We attribute this small positive MR to a angular mismatch (see below). It should be pointed out that the sample was installed on a home-built sample stage in which an angular mismatch (both polar and azimuthal) up to $\sim$5 \textordmasculine could be possible with respect to a veritable LMR. The situation is similar when all the electrodes were fully painted across the width ($R_E$, Figure S4c), but the angular mismatch seems even larger than the previous measurements. Note that these positive MR regions completely disappear when the azimuthal magnetic fields ($B_x$ and $B_y$) with proper values are turned on to overcome the angular mismatch, see the green curve ($R_{E'}$) in Figure S4c. This field refinement also suppresses the small upturn at high field as mentioned in the maintext. It is interesting to compare the field dependence of $R_{E'}$ to $R_B$. One clearly sees that both show negative MR under field and saturate to comparable values at high field, but $R_B$ drops much faster that $R_{E'}$. All these measurements manifest an intrinsic negative LMR in TaAs$_2$, although the current-jetting effect could play some role if contacts are not properly made.

\end{document}